\documentclass[aps,pre,twocolumn,showpacs,superscriptaddress,10pt]{revtex4} 
\usepackage{graphicx}
\usepackage{subfigure}
\usepackage{amsmath}
\usepackage{amsfonts}
\usepackage{amssymb}
\usepackage{amsthm}
\usepackage{times}
\usepackage[colorlinks,citecolor=blue,linkcolor=blue]{hyperref}
\usepackage{color}
\usepackage{setspace}

\begin{document}
\title{Frequency-dependent current noise in quantum heat transfer with full counting statistics}

\newcommand{\MIT}{\affiliation{Department of Chemistry, Massachusetts Institute of Technology, 77 Massachusetts Avenue, Cambridge, MA 02139, USA}}
\newcommand{\Smart}{\affiliation{Singapore-MIT Alliance for Research and Technology (SMART) center, 1 CREATE Way, Singapore 138602, Singapore}}
\newcommand{\Fudan}{\affiliation{State Key Laboratory of Surface
Physics and Department of Physics, Fudan University, Shanghai 200433,
China}}

\author{Junjie Liu}
\Smart
\MIT
\author{Chang-Yu Hsieh}
\Smart
\MIT
\author{Changqin Wu}
\Fudan
\author{Jianshu Cao}
%\email{jianshu@mit.edu}
\MIT
\Smart

\newcommand{\FD}{FDCN}

\begin{abstract}
To investigate frequency-dependent current noise (\FD) in open quantum systems at steady states, we present a theory which combines Markovian quantum master equations with a finite time full counting statistics. Our formulation of the \FD{} generalizes previous zero-frequency expressions and can be viewed as an application of MacDonald's formula for electron transport to heat transfer. As a demonstration, we consider the paradigmatic example of quantum heat transfer in the context of a non-equilibrium spin-boson model.  We adopt a recently developed polaron-transformed Redfield equation which allows us to accurately investigate heat transfer with arbitrary system-reservoir coupling strength, arbitrary values of spin bias as well as temperature differences. We observe maximal values of \FD{} in moderate coupling regimes, similar to the zero-frequency cases. We find the \FD{} with varying coupling strengths or bias displays a universal Lorentzian-shape scaling form in the weak coupling regime, and a white noise spectrum emerges with zero bias in the strong coupling regime due to a distinctive spin dynamics. We also find the bias can suppress the \FD{} in the strong coupling regime, in contrast to its zero-frequency counterpart which is insensitive to bias changes. Furthermore, we utilize the Saito-Utsumi relation as a benchmark to validate our theory and study the impact of temperature differences at finite frequencies. Together, our results provide detailed dissections of the finite time fluctuation of heat current in open quantum systems. 
\end{abstract}

%05.60.Gg:quantum transport
%05.30.-d:quantum statistical mechanics
%05.40.-a:fluctuation phenomena, noise, random processes
%44.10.+i:heat conduction

\maketitle

\section{Introduction}
The rapid development in nanotechnologies opens an avenue for studying heat transfer in mesoscopic systems \cite{Lee.13.N,Jezouin.13.S}. At the nano-scales, fluctuations of heat become increasingly relevant \cite{Battista.13.PRL} to the perfomance and stability of the nanostructured devices. To better characterize the fluctuations, higher order statistics of heat transfer beyond stationary heat current are needed and cannot be directly obtained from the standard heat conductance measurements. Hence, it is desirable to formulate a theoretical framework to analyze heat-transfer statistics for these systems.

So far, analytical results on the heat-transfer statistics are limited to the infinite time limit (i.e. zero frequency limit) where well-established frameworks such as the large deviation theory \cite{Touchette.09.PR}, steady state fluctuation theorem \cite{Jarzynski.04.PRL,Esposito.09.RMP,Seifert.12.RPP} and full counting statistics (FCS) \cite{Levitov.93.PZ,Bagrets.03.PRB,Agarwalla.12.PRE} can be utilized. Both the stationary heat current \cite{Boudjada.14.JPCA,Wang.15.SR,Liu.17.PRE} and variance of heat current have been studied for open quantum systems at steady states \cite{Saito.07.PRL,Nicolin.11.JCP,Nicolin.11.PRB,Wang.17.PRA,Bijay.17.NJP}. However, this is by no means the complete story. Finite-frequency components of heat fluctuations provides a rich set of new information about the steady state heat statistics beyond what could be inferred from the zero-frequency component, as already demonstrated for electronic heat transport in the wide-band limit \cite{Zhan.11.PRB}. Apart from this example, and some notable exceptions \cite{Krive.01.PRB,Averin.10.PRL}, the behaviors of the finite frequency heat-transfer statistics, is still largely unexplored due to the absence of a general theoretical framework that can extract finite time fluctuation properties of heat at steady states.

In this work, we present a theoretical method to study the frequency-dependent current noise (\FD) for nonequilibrium open quantum systems \cite{Breuer.07.NULL,Weiss.12.NULL,Schaller.14.NULL}. We extend MacDonald's formula \cite{MacDonald.49.RPP,Lambert.07.PRB} in electron transport to heat current which formally expresses the \FD{} in terms of an integral of the time-dependent second order cumulant of transferred heat evaluated at steady states, thus generalizing previously expressions for zero-frequency heat current noise \cite{Nicolin.11.JCP,Nicolin.11.PRB,Wang.17.PRA,Bijay.17.NJP}. In order to calculate the time-dependent cumulant of heat involved in the \FD{}, we follow the scheme of a finite time FCS developed for electron transport \cite{Marcos.10.NJP} and propose an analogous framework. Our theory can be applied to open quantum systems described by Markovian quantum master equations and possesses good adaptability. We remind that the finite-time situation considered in this work is
not the same as the transient response\cite{Cerrillo.16.PRB} of an open quantum system with some prescribed initial conditions. Fig. \ref{fig:counting} clearly summarizes how the finite-time heat fluctuation should be understood in the present work.

To illustrate the formalism, we study the case of a nonequilibrium spin-boson (NESB) model \cite{Leggett.87.RMP,Weiss.12.NULL} which is a paradigmatic example for quantum heat transfer \cite{Boudjada.14.JPCA}. By combining a recently developed nonequilibrium polaron-transformed Redfield equation (NE-PTRE) for the reduced spin dynamics \cite{Wang.15.SR,Wang.17.PRA} and the finite time FCS, we are able to study the \FD{} of the NESB from a unified perspective. New phenomena are found and explained analytically. These results manifest the versatility of our proposed framework and how it can be used to study \FD{} in variety of open
quantum system contexts\cite{Wang.14.NJP,Xu.16.NJP,Thingna.16.SR}.
%which make our theory a promising method to address the FD-HCNP in quantum open systems.

The paper is organized as follows. We first propose our general theory for the \FD{} in section \ref{sec:2}. In section \ref{sec:3}, we introduce the NESB model and the NE-PTRE with FCS. In section \ref{sec:4}, we study the \FD{} of the NESB in detail by analyzing the impact of coupling strength, bias as well as temperature differences. In section \ref{sec:5}, we summarize our findings.

\section{Theory}\label{sec:2}
\subsection{Frequency-dependent heat current noise power}
We consider heat transfer systems, consisting of a central region attached to non-interacting bosonic reservoirs at {\it different temperatures}. This setup can be described by a general Hamiltonian
\begin{equation}\label{eq:H_general}
H ~=~ H_s+H_I+H_B,
\end{equation}
where $H_s$ refers to the system, $H_B=\sum_{v=L,R}H_B^v=\sum_{k,v=L,R}\omega_{k,v}b^{\dagger}_{k,v}b_{k,v}$ is the bosonic reservoirs part with $b_{k,v}^{\dagger}$ and $b_{k,v}$ the bosonic creation and annihilation operators for the mode $k$ of frequency $\omega_{k,v}$ in the $v$-th reservoir characterized by an inverse temperature $\beta_v\equiv T_v^{-1}$ ($T_L\neq T_R$),
$H_I=\sum_vV_v\otimes B_v$ is the interaction between the system and reservoirs which assumes a bilinear form with $B_v$ an arbitrary operator of $v$-th reservoir and $V_v$ the corresponding system operator. This setup encompasses a broad range of dissipative and transport settings. Throughout the paper, we set $\hbar=1$ and $k_B=1$.  

The stationary heat current is defined by
\begin{equation}\label{eq:hc}
\langle I_v(t)\rangle~=~-\frac{d}{dt}\langle H_B^v(t)\rangle,
\end{equation}
where we denote $\langle\cdots\rangle\equiv\mathrm{Tr}[\cdots\rho_{ss}]$ with $\rho_{ss}$ the total steady state density matrix. Due to the energy conservation, we introduce $\langle I\rangle\equiv\langle I_L\rangle=-\langle I_R\rangle$. We simply focus on the heat current $I$ and its fluctuation statistics. It is worthwhile to mention that the above definition is consistent with the quantum thermodynamics and can be applied in the strong coupling regime \cite{Esposito.15.PRL}.

We assume the total system has reached the unique steady state at $t=0$, then the heat current noise at finite times is described by the symmetrized auto-correlation function \cite{Zhan.11.PRB}
\begin{equation}\label{eq:c}
S(t_1,t_2)~=~\frac{1}{2}\langle\{\Delta I(t_1),\Delta I(t_2)\}\rangle
\end{equation}
with $\Delta I(s)=I(s)-\langle I\rangle$ the fluctuation of time dependent heat current $I(s)=-\frac{d}{dt}H_B^L(s)$ from its average value, where the anti-commutator $\{A,B\}=AB+BA$ ensures the Hermitian property. At steady states, the correlation function only depends on the time difference such that $S(t_1,t_2)=S(\tau)$ with $\tau=|t_1-t_2|$ the time interval. Therefore, the Fourier transform yields the \FD{} $S(\omega)$ for the heat current
\begin{equation}
S(\omega)~=~S(-\omega)=\int_{-\infty}^{\infty}d\tau e^{i\omega \tau}S(\tau) \geq 0.
\end{equation} 
Since $S(\omega)$ is an even function in frequency and strictly semi-positive in accordance with the Wiener-Khintchine theorem, in the following we consider positive values of the frequency, $\omega>0$, only.

According to the definition of heat current in Eq. (\ref{eq:hc}), we can introduce $Q(t)=H_B^L(t)-H_B^L(0)$ as the heat transferred from the left reservoir to the right reservoir in the time span $0$ to $t$ (we assume the left reservoir has a higher temperature) with $\langle Q(t)\rangle=\langle I\rangle t$, and in analogy with MacDonald's formula in electron charge transport \cite{MacDonald.49.RPP,Lambert.07.PRB}, we find
\begin{equation}\label{eq:swf}
S(\omega)~=~\omega\int_0^{\infty}dt\sin(\omega t)\frac{\partial}{\partial t}\langle Q^2(t)\rangle_c,
\end{equation}
where we define the second order cumulant of heat as $\langle Q^2(t)\rangle_c\equiv\langle Q^2(t)\rangle-\langle Q(t)\rangle^2$. Eq. (\ref{eq:swf}) can be viewed as an application of the MacDonald's formula in electron charge transport to heat transfer. However, in contrast to the electron charge current, the above MacDonald-like formula represents the total heat current noise spectrum due to the absence of a displacement current component in heat transfer setups. This formally exact relation enables us to calculate \FD{} from the finite time heat statistics. The second order cumulant involved in the above definition can be obtained from the cumulant generating function (CGF) which is the main focus of the FCS ( see, e.g., Ref. \cite{Esposito.09.RMP} and reference therein). However, instead of considering the FCS in the infinite time limit as in previous studies , we should follow the framework of a finite time FCS \cite{Marcos.10.NJP} such that finite time properties of cumulants which are essential for finite-frequency noise power can be extracted.

The above finite frequency definition for $S(\omega)$ can recover the well-known zero-frequency expression. To see this, we introduce the regularization \cite{Flindt.05.PE}
\begin{equation}
\omega\sin\omega t ~=~ (\omega\sin\omega t+\varepsilon\cos\omega t)e^{-\varepsilon t},~~\varepsilon\rightarrow 0^{+},
\end{equation}
which ensures correct results at $\omega= 0$ case. Then we find from Eq. (\ref{eq:swf}) that
\begin{equation}
S(0)~=~\left.\varepsilon\int_0^{\infty}e^{-\varepsilon t}\frac{\partial}{\partial t}\langle Q^2(t)\rangle_c\right|_{\varepsilon\rightarrow 0^{+}}.
\end{equation}
By using the final value theorem of the Laplace transform, we obtain
\begin{equation}\label{eq:s0}
S(0) ~=~ \left.\frac{\partial}{\partial t}\langle Q^2(t)\rangle_c\right|_{t\to\infty}.
\end{equation}
In the infinite time limit (see Fig. \ref{fig:counting}), all cumulants increase linearly in time $t$ as guaranteed by the FCS \cite{Esposito.09.RMP}, then the above relation is just the expression utilized in recent studies on $S(0)$ \cite{Nicolin.11.JCP,Nicolin.11.PRB,Wang.17.PRA,Bijay.17.NJP}. Therefore, Eq. (\ref{eq:swf}) generalize previous zero-frequency expressions.

\subsection{Finite time full counting statistics}
\subsubsection{Finite time generating functions}
\begin{figure}[tbh!]
  \centering
  \includegraphics[width=1\columnwidth]{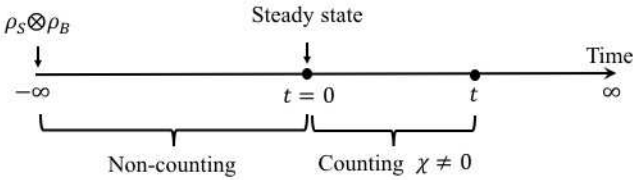}
\caption{(Color online) Illustration of \textquotedblleft finite time" parameter regime. The total system has reached the steady state at $t=0$. The two-time measurements are taken at $t=0$ and $t$ with a nonzero counting field $\chi$.}
\label{fig:counting}
\end{figure}

In accordance with the definition of heat current, we study the statistics of heat $Q(t)$ transferred from the left reservoir to the right reservoir during a time interval $[0,t]$. The specific measurement of the net transferred heat $Q(t)$ is performed using a two-time measurement protocol \cite{Esposito.09.RMP,Campisi.11.RMP}: Initially at time $t=0$ where the total system has reached the steady state, we introduce a projector $K_{q_0}=|q_0\rangle\langle q_0|$ with $|q_0\rangle$ one of eigen-states of the left bath Hamiltonian $H_B^L$ to measure the quantity $H_B^L$, giving an outcome $q_0$. A second measurement is performed at time $t$ with a projector $K_{q_t}=|q_t\rangle\langle q_t|$ ($|q_t\rangle$ is also one of eigen-states of the left bath Hamiltonian $H_B^L$) and an outcome $q_t$. Hence, the measurement outcome of net transferred heat is determined by $Q(t)=q_t-q_0$. The corresponding joint probability to measure $q_0$ at $t=0$ and $q_t$ at time $t$ reads 
\begin{equation}
P[q_t,q_0]~\equiv~\mathrm{Tr}\{K_{q_t}U(t,0)K_{q_0}\rho(0)K_{q_0}U^{\dagger}(t,0)K_{q_t}\},
\end{equation}
where $U(t,0)$ the unitary time evolution operator of the total system, $\rho(0)$ is the total density matrix when the counting starts. We should choose $\rho(0)=\rho_{ss}$ as we are interested in fluctuations at steady states. Such an initial condition can be constructed by switching on the interaction $H_I$ from the infinite past, where the density matrix $\rho(-\infty)$ is given by a direct product of density matrices of system $\rho_s$ and bosonic baths $\rho_B$. The corresponding counting scheme is illustrated in Fig. \ref{fig:counting}.
The probability distribution for the difference $Q(t)=q_t-q_0$ between the output of the two measurement is given by
\begin{equation}\label{eq:pqt}
p(Q,t)~=~\sum_{q_t,q_0}\delta(Q(t)-(q_t-q_0))P[q_t,q_0],
\end{equation} 
where $\delta(x)$ denotes the Dirac distribution. Then we can introduce a finite time moment generating function (MGF) associated with this probability
\begin{equation}\label{eq:mgf}
Z(\chi,t)=\int\,dQ(t) p(Q,t)e^{i\chi Q(t)}
\end{equation}
with $\chi$ the counting-field parameter. Its logarithm gives the CGF $G(\chi,t)=\ln Z(\chi,t)$.

To formulate explicit expressions for generating functions that are suitable for analytical as well as numerical studies. We focus on open quantum systems described by a reduced density matrix $\rho_s(t)$ which obeys a generalized Markovian master equation 
\begin{equation}\label{eq:gme}
\dot{\rho}_s(t)~=~-L\rho_s(t)
\end{equation}
with $\dot{A}(t)\equiv\frac{\partial A(t)}{\partial t}$ and $L$ the Liouvillian operator driving the dynamics of the system. Although we limit ourselves to Markovian master equations, an extension to include non-Markovian effects \cite{Braggio.06.PRL,Flindt.08.PRL} is possible which will be addressed in the future work.

In order to investigate the statistics of transferred heat during the time span $[0,t]$ at steady states, we proceed by projecting the reduced density matrix $\rho_s(t)$ onto the subspace of $Q$ net transferred heat, and denote this $Q$-resolved density matrix as $\rho_s(Q,t)$, in analogy with $n$-resolved density matrix in quantum optics \cite{Cook.81.PRA} and mesoscopic electron transport \cite{Gurvitz.98.PRB}. The probability $p(Q,t)$ in Eq. (\ref{eq:pqt}) is just
\begin{equation}
p(Q,t)~=~\mathrm{Tr}[\rho_s(Q,t)].
\end{equation}
The MGF can conveniently be expressed as
\begin{equation}\label{eq:z1}
Z(\chi,t)~=~\mathrm{Tr}[\rho_s(\chi,t)]
\end{equation}
in terms of the $\chi$-dependent density matrix $\rho_s(\chi,t)=\int dQ e^{i\chi Q}\rho_s(Q,t)$. We note that by setting $\chi=0$, we recover the original density matrix: $\rho_s(\chi=0,t)=\rho_s(t)$. 

To evaluate the MGF and CGF we consider a modified master equation governing the evolution of the $\chi$-dependent density matrix. According to Eq. (\ref{eq:gme}), the modified master equation takes the form
\begin{equation}\label{eq:rhoc}
\dot{\rho}_s(\chi,t)~=~-L_{\chi}\rho_s(\chi,t)
\end{equation}
with a $\chi$-dependent Liouvillian $L_{\chi}$. Since Eq. (\ref{eq:rhoc}) is a linear differential equation for $\rho_s(\chi,t)$, it can be rewritten in the form
\begin{equation}
|\dot{\rho}_s(\chi,t)\rangle\rangle~=~-\mathbb{L}_{\chi}|\rho_s(\chi,t)\rangle\rangle,
\end{equation}
where $|\rho_s(\chi,t)\rangle\rangle$ is the vector representation of $\rho_s(\chi,t)$ and $\mathbb{L}_{\chi}$ is the matrix representation of $L_{\chi}$ in the Liouville space \cite{Esposito.09.RMP} (Here we use double angle brackets to distinguish these vectors from the ordinary quantum mechanical "bras" and "kets" ). By formally solving the above equation we find
\begin{equation}\label{eq:z2}
|\rho_s(\chi,t)\rangle\rangle~=~e^{-\mathbb{L}_{\chi}t}|\rho_{s}^{stat}\rangle\rangle,
\end{equation}
as we require that the system at $t=0$ has reached the steady state defined by $\mathbb{L}|\rho_{s}^{stat}\rangle\rangle=0$ or by simply tracing out the reservoir degrees of freedom in the total steady state $\rho_{ss}$. Since the Liouvillian $L$ conserves probability, it holds that $\mathrm{Tr}[L\rho_s(t)] = 0$ for any density matrix. This implies that the left zero-eigenvector of $\mathbb{L}$ is the vector representation of the trace operation, i.e., $\langle\langle \tilde{0}|\mathbb{L}=0$ and $\langle\langle \tilde{0}|\rho_{s}^{stat}\rangle\rangle=\mathrm{Tr}[\rho_{s}^{stat}]=1$. Combining Eqs. (\ref{eq:z1}) and (\ref{eq:z2}), we then obtain the following compact expression:
\begin{equation}\label{eq:z3}
Z(\chi,t)~=~\langle\langle \tilde{0}|e^{-\mathbb{L}_{\chi}t}|\rho_{s}^{stat}\rangle\rangle\equiv \langle\langle e^{-\mathbb{L}_{\chi}t}\rangle\rangle.
\end{equation}
This expression holds for a system that has been prepared in an arbitrary state in the far past. The system has then evolved until $t = 0$, where it has reached the steady state. At $t = 0$, we start collecting statistics to construct the probability distribution $p(Q,t)$ for $Q$ net transferred heat in the time span $[0,t]$. 

By diagonalizing the $\mathbb{L}_{\chi}$ as
\begin{equation}
\mathbb{L}_{\chi}~=~\sum_n\mu_n(\chi)|f_n(\chi)\rangle\rangle\langle\langle g_n(\chi)|
\end{equation}
with $\mu_n$ the $n$-th eigenvalue, $|f_n(\chi)\rangle\rangle$ and $\langle\langle g_n(\chi)|$ the corresponding right and left eigenvectors, respectively. In the $\chi\to 0$ limit, one of these eigenvalues, $\mu_0(\chi)$ say, tends to zero and the corresponding eigenvectors gives the stationary states $\langle\langle \tilde{0}|$ and $|\rho_{s}^{stat}\rangle\rangle$ for the system. This single eigenvalue is sufficient to determine the zero-frequency FCS \cite{Bagrets.03.PRB}. In contrast, here we need all eigenvalues and eigenvector in constructing finite time FCS. We can rewrite the MGF as
\begin{equation}\label{eq:z_e}
Z(\chi,t)~=~\sum_ne^{-\mu_n(\chi)t}F_n(\chi)
\end{equation}
with $F_n(\chi)=\langle\langle \tilde{0}|f_n(\chi)\rangle\rangle\langle\langle g_n(\chi)|\rho_s^{stat}\rangle\rangle$. Consequently, the CGF can be expressed as $G(\chi,t)=\ln\sum_ne^{-\mu_n(\chi)t}F_n(\chi)$.

\subsubsection{Expressions of \FD{}}
Introducing the $n$th cumulant $\langle Q^n(t)\rangle_c$ of $Q(t)$ as
\begin{equation}
\langle Q^n(t)\rangle_c~=~\left.\frac{\partial^{n}}{\partial(i\chi)^n}G(\chi,t)\right|_{\chi=0},
\end{equation}
and the coefficients
\begin{equation}\label{eq:gsu}
C_m^n(t) ~=~\left.\frac{\partial^{n+m}}{\partial(i\chi)^n\partial A^m}G(\chi,t)\right|_{\chi=0},
\end{equation}
where $A\equiv\beta_R-\beta_L$ denotes the affinity, we will have
\begin{equation}\label{eq:sw_formal}
S(\omega) ~=~ \omega\int_0^{\infty}dt\sin\omega t\frac{\partial}{\partial t}C_0^2(t).
\end{equation}
This formal exact relation represents a nonlinear superposition of thermal (Johson-Nyquist), shot and quantum noises and enables us to investigate the \FD{} within the framework of finite time FCS for open quantum systems, thus constituting one of main results in this work. If we introduce $J_m^n$ as the long time limit of $\frac{\partial}{\partial t}C_m^n(t)$ then $S(0)=J_0^2$ according to Eq. (\ref{eq:s0}). 

Eq. (\ref{eq:sw_formal}) is particularly useful in obtaining analytical results provided that the dimension of the Liouvillian operator $\mathbb{L}_{\chi}$ is small. If the eigen-problem of the Liouvillian operator becomes complicated, we should resort to a numerical treatment. By noting 
\begin{equation}
C^2_0(t)~=~\langle Q^2(t)\rangle-\langle Q(t)\rangle^2=\langle Q^2(t)\rangle-(It)^2
\end{equation} 
in steady states with $\langle Q^2(t)\rangle$ the second order moment of heat $Q(t)$ generated by the MGF: $\left.\frac{\partial^2}{\partial(i\chi)^2}Z(\chi,t)\right|_{\chi=0}$. Inserting this expression into Eq. (\ref{eq:sw_formal}) and performing the integral, we have
\begin{equation}
S(\omega)~=~\omega\frac{\partial^2}{\partial(i\chi)^2}\int_0^{\infty}\,dt\sin\omega t\frac{\partial}{\partial t}\left.Z(\chi,t)\right|_{\chi=0}.
\end{equation}
Utilizing the Laplace transform ($t\to \lambda$), the above equation can be cast into \cite{Flindt.08.PRL}
\begin{equation}\label{eq:sw_SB}
S(\omega)~=~-\frac{\omega^2}{2}\frac{\partial^2}{\partial(i\chi)^2}\left.\left[\langle\langle\Omega(\chi,\lambda=i\omega)+\Omega(\chi,\lambda=-i\omega)\rangle\rangle\right]\right|_{\chi=0},
\end{equation}
where we denote $Z(\chi,\lambda)=\langle\langle\Omega(\chi,\lambda)\rangle\rangle$ with $\Omega(\chi,\lambda)=(\lambda+\mathbb{L}_{\chi})^{-1}$. This expression is more suitable for numerical simulations as only the stationary state at $t=0$ and Liouvillian operator are needed.
 
\section{Nonequilibrium spin-boson model}\label{sec:3}
\subsection{Model setup}
To illustrate our general method, we analyze the statistics of the prototypical example of quantum heat transfer through a NESB model \cite{Leggett.87.RMP,Weiss.12.NULL}. The NESB consists of a two-level spin in the central region and is described by the Hamiltonian 
\begin{equation}\label{eq:hs}
H_s ~=~ \frac{\varepsilon_0}{2}\sigma_z+\frac{\Delta}{2}\sigma_x,
\end{equation}
where $\varepsilon_0$ is the bias, $\Delta$ is the tunnelling between two levels and $\sigma_{x,z}$ are the Pauli matrices. Since the spectrum of the system Hamiltonian is a symmetric function of bias, here we only consider positive bias. The operators in the interaction term $H_I$ read
\begin{equation}\label{eq:bv}
V_v ~=~ \sigma_z,~~~B_v=\sum_k g_{k,v}(b^{\dagger}_{k,v}+b_{k,v})
\end{equation}
with $g_{k,v}$ the system-reservoir coupling strength. The influence of bosonic reservoirs are characterized by a spectral density $\gamma_v(\omega)=2\pi\sum_kg_{k,v}^2\delta(\omega-\omega_{k,v})$. For reservoirs with infinite degrees of freedom, $\gamma_v(\omega)$ can be regarded as a continuous function of its argument, then we can let $\gamma_v(\omega)=\pi\alpha_v\omega^s\omega_{c,v}^{1-s}e^{-\omega/\omega_{c,v}}$ with $\alpha_v$ the dimensionless system-reservoir coupling strength of the order of $g_{k,v}^2$ and $\omega_{c,v}$ the cut-off frequency of the $v$-th bosonic reservoir. For simplicity and without loss of generality, we consider the super-Ohmic spectrum $s=3$ which is of experimental relevance \cite{Goerlich.89.EL}, and choose $\alpha_L=\alpha_R=\alpha$, $\omega_{c,L}=\omega_{c,R}=\omega_c$.

We limit our calculation to the so-called nonadiabatic limit of $\Delta/\omega_c\ll1$. For fast reservoirs, it has been demonstrated that the polaron transformation (PT) is suitable for the entire range of system-bath coupling strength \cite{Lee.12.JCP,Wang.15.SR,Liu.16.CP,Liu.17.PRE} and enables us to study the impact of system-reservoir interaction beyond the weak coupling limit. Thus we perform the PT with the unitary operator
\begin{equation}\label{eq:U}
U=\mathrm{exp}[i\sigma_z\Phi/2],\qquad\Phi=2i\sum_{k,v}\frac{g_{k,v}}{\omega_{k,v}}(b^{\dagger}_{k,v}-b_{k,v})
\end{equation}
such that
\begin{equation}
H_T ~=~ U^{\dagger}HU= \tilde{H}_0+\tilde{H}_I,
\end{equation}
where the free Hamiltonian is $\tilde{H}_0=\tilde{H}_s+\tilde{H}_B$ with the reservoir Hamiltonian remains unaffected, $\tilde{H}_B=H_B$, and the transformed system Hamiltonian reads
\begin{equation}
\tilde{H}_s=\frac{\varepsilon_0}{2}\sigma_z+\frac{\eta\Delta}{2}\sigma_x,
\end{equation}
where the renormalization factor due to the formation of polarons reads \cite{Segal.06.PRB,Chen.13.PRB,Wang.15.SR,Wang.17.PRA}
\begin{equation}
\eta~=~\exp\left(-\sum_v\int_0^{\infty}d\omega\frac{\gamma_v(\omega)}{2\pi\omega^2}\coth\frac{\beta_v\omega}{2}\right).
\end{equation}
For the super-Ohmic spectrum $s=3$ we consider here, the renormalization factor is specified as $\eta=\exp\{-\sum_v\alpha[-1+\frac{2}{(\beta_v\omega_c)^2}\psi_1(1/\beta_v\omega_c)]/2\}$, with the trigamma function $\psi_1(x)=\sum_{n=0}^{\infty}1/(n+x)^2$. As can be seen, in the weak coupling regime, $\eta$ becomes $1$, while in the strong coupling regime, it vanishes. The transformed interaction term, originated from the tunneling term in Eq. (\ref{eq:hs}), takes the following form
\begin{equation}\label{eq:t_interaction}
\tilde{H}_I~=~\frac{\Delta}{2}[\sigma_x(\cos\Phi-\eta)+\sigma_y\sin\Phi].
\end{equation}
It's evident that $\tilde{H}_I$ contains arbitrary orders of the system-reservoir coupling strength by noting the form of $\Phi$ in Eq. (\ref{eq:U}), however, its average vanishes. Hence we can treat $\tilde{H}_I$ perturbatively regardless of coupling strength. Therefore, in the polaron picture, we can study finite time FCS from the weak to strong system-reservoir coupling regime.

\subsection{Nonequilibrium polaron-transformed Redfield equation}
In order to study finite time FCS of heat in the polaron picture, we follow a recently developed NE-PTRE method which leads to the following master equation for the reduced density matrix $\rho_s(t)$ \cite{Wang.15.SR,Wang.17.PRA}
\begin{eqnarray}\label{eq:ptre}
\dot{\rho}_s(t) &=& -i[\tilde{H}_s,\rho_s]+\sum_{l=e,o}\sum_{\omega,\omega^{\prime}=0,\pm\omega_0}\Gamma_l(\omega)[P_l(\omega)\rho_s,P_l(\omega^{\prime})]\nonumber\\
&&+\mathrm{H.c.}, 
\end{eqnarray} 
where $\omega_0=\sqrt{\varepsilon_0^2+\eta^2\Delta^2}$ is the energy gap in the eigenbasis, $P_{e(o)}(\omega)$ is the transition projector in the eigenbasis obtained from the evolution of Pauli matrices $\sigma_{x(y)}(-\tau)=\sum_{\omega=0,\pm\omega_0}P_{e(o)}e^{i\omega\tau}$, The subscript $e(o)$ denotes the even (odd) parity of transfer dynamics. The transition rates are $\Gamma_o(\omega)=\left(\frac{\eta\Delta}{2}\right)^2\int_0^{\infty}d\tau e^{i\omega\tau}\sinh[Q(\tau)]$ and $\Gamma_e(\omega)=\left(\frac{\eta\Delta}{2}\right)^2\int_0^{\infty}d\tau e^{i\omega\tau}(\cosh[Q(\tau)]-1)$ with
$Q(\tau)$ denotes the sum of bosonic correlation functions $Q(\tau)=\sum_vQ_v(\tau)$: 
\begin{equation}
Q_v(\tau)~=~\int_0^{\infty}d\omega\frac{\gamma_v(\omega)}{\pi\omega^2}\left[\coth\frac{\beta_v\omega}{2}\cos\omega\tau-i\sin\omega\tau\right].
\end{equation}
As clearly demonstrated in Ref. \cite{Wang.17.PRA}, $\Gamma_{o(e)}(\omega)$ describe totally different transfer processes. %we omit the discussion here for simplicity.

Combining Eq. (\ref{eq:ptre}) with counting field $\chi$, we have following $\chi$-dependent NE-PTRE for the $\chi$-dependent density matrix \cite{Wang.17.PRA}
\begin{equation}\label{eq:ptre_chi}
\dot{\rho}_s(\chi,t)~=~ -i[\tilde{H}_s,\rho_s(\chi,t)]+\mathcal{D}_{\chi}[\rho_s(\chi,t)],
\end{equation} 
where the dissipator reads $\mathcal{D}_{\chi}[\rho_s]=\sum_{l}\sum_{\omega,\omega^{\prime}}\{[\Gamma_{l,-}^{\chi}(\omega)+\Gamma_{l,+}^{\chi}(\omega^{\prime})]
P_l(\omega^{\prime})\rho_sP_l(\omega)-[\Gamma_{l,+}(\omega)P_l(\omega^{\prime})P_l(\omega)\rho_s+\mathrm{H.c.}]\}$
and $\chi$-dependent transition rates are expressed as ($\sigma=\pm$)
\begin{eqnarray}
&&\Gamma_{e,\sigma}^{\chi}(\omega) = \left(\frac{\eta\Delta}{2}\right)^2\int_0^{\infty}d\tau e^{i\omega\tau}[\cosh Q(\tau_{\sigma}^{\chi})-1],\nonumber\\
&&\Gamma_{o,\sigma}^{\chi}(\omega) = \left(\frac{\eta\Delta}{2}\right)^2\int_0^{\infty}d\tau e^{i\omega\tau}\sinh Q(\tau_{\sigma}^{\chi}),
\end{eqnarray}
where $\tau_{\sigma}^{\chi}\equiv\sigma\tau-\chi$ and $\chi$-dependent bosonic correlation function becomes $Q(\tau-\chi)=Q_L(\tau-\chi)+Q_R(\tau)$. 

Defining the vector form of the $\chi$-dependent reduced density matrix as $|\rho_s(\chi,t)\rangle\rangle=[P_{11}^{\chi},P_{00}^{\chi},P_{10}^{\chi},P_{01}^{\chi}]^T$ with $P_{ij}^{\chi}=\langle i|\rho_s(\chi,t)|j\rangle$, we can express the $\chi$-dependent NE-PTRE in the Liouville space, i.e., $|\dot{\rho}_s(\chi,t)\rangle\rangle=-\mathbb{L}_{\chi}|\rho_s(\chi,t)\rangle\rangle$. 
%For instance, $\mathbb{L}_{\chi}$ for systems without bias has the following explicit form \cite{Wang.17.PRA}
%\begin{equation}
%\mathbb{L}_{\chi}~=~\left(\begin{array}{cccc}
%a & -a_{\chi} & b_{\chi} & c_{\chi}\\
%-a_{\chi} & a & c_{\chi} & b_{\chi}\\
%d_{\chi} & e_{\chi} & a & f_{\chi}\\
%e_{\chi} & d_{\chi} & f_{\chi} & a
%\end{array}\right).
%\end{equation}
%The matrix elements are defined as $a_{\chi}=X_{e}^{\chi}+Y_{\chi}/2$, $b_{\chi}=-(X_{o,+}^{\chi}+X_{o,-})/2$, $c_{\chi}=(X_{o,+}+X_{o,-}^{\chi})/2$, $d_{\chi}=(X_{o,+}^{\chi}-X_{o,-})/2$, $e_{\chi}=(X_{o,+}-X_{o,-}^{\chi})/2$, $f_{\chi}=-X_{e}^{\chi}+Y_{\chi}/2$ and $a=a_{\chi=0}$, with the modified transition rates $X_{e}^{\chi}=\Gamma_{e,+}^{\chi}(0)+\Gamma_{e,-}^{\chi}(0)$, $Y_{\chi}=\Gamma_{o,+}^{\chi}(\omega_0)+\Gamma_{o,+}^{\chi}(-\omega_0)+\Gamma_{o,-}^{\chi}(\omega_0)+\Gamma_{o,-}^{\chi}(-\omega_0)$ and $X_{o,\pm}^{\chi}=\Gamma_{o,\pm}^{\chi}(\omega_0)-\Gamma_{o,\pm}^{\chi}(-\omega_0)$. 
Inserting the form of $\mathbb{L}_{\chi}$ obtained from the $\chi$-dependent NE-PTRE Eq. (\ref{eq:ptre_chi})  into the formal expression for $S(\omega)$ Eq. (\ref{eq:sw_SB}), we can obtain explicit numerical results for the \FD{} as a function of coupling strength, bias and temperature differences. In the following, we will provide detailed results.

\section{Results}\label{sec:4}
\subsection{Effect of coupling strength}
\begin{figure}[tbh!]
  \centering
  \includegraphics[width=1\columnwidth]{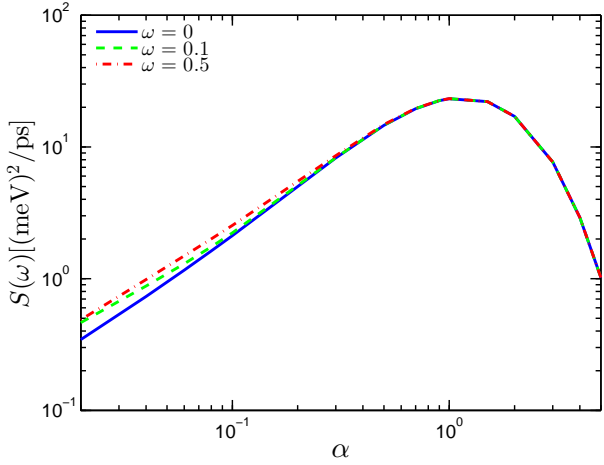}
\caption{(Color online) Behaviors of noise spectrum $S(\omega)$ with varying system-reservoir coupling strength $\alpha$ for different frequencies. Other parameters are $\Delta=5.22$meV, $\omega_c=26.1$meV, $\varepsilon_0=0$, $T_L=180$K, $T_R=90$K.}
\label{fig:Sw_alpha}
\end{figure}

\begin{figure}[tbh!]
  \centering
  \includegraphics[width=1\columnwidth]{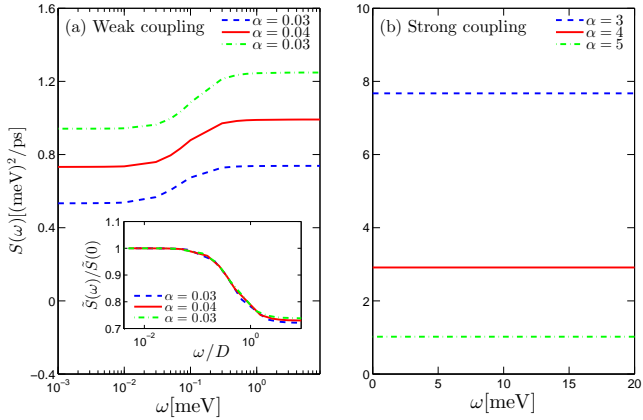}
\caption{(Color online) Behaviors of noise spectrum $S(\omega)$ as a function of $\omega$ with varying coupling strength $\alpha$ for (a) weak couplings and (b) strong couplings. The inset in (a) shows universal behaviors of a scaled noise power $\tilde{S}(\omega)$ [defined in Eq. (\ref{eq:ssw})] in the weak coupling regime., the scale parameter $D$ is the sum of total relaxation and activation rates in the Redfield picture. Other parameters are $\Delta=5.22$meV, $\omega_c=26.1$meV, $\varepsilon_0=0$meV, $T_L=180$K, $T_R=90$K.}
\label{fig:Sw_a}
\end{figure}

We first investigate the behaivors of $S(\omega)$ with varying system-reservoir coupling strength under the condition of fixed temperatures and zero bias. Typical numerical results are shown in the Fig. \ref{fig:Sw_alpha}. We find that even at finite frequencies, the noise spectrum still depicts a non-monotonic turnover behavior in the intermediate coupling regime as that in the zero-frequency case \cite{Wang.17.PRA}.
Another interesting finding is that $S(\omega)$ has distinct frequency dependences in the weak and strong coupling regimes (details are listed in Fig. \ref{fig:Sw_a}): In the weak coupling regime, $S(\omega)$ is a monotonic increasing function of $\omega$ and saturates at high frequency [Fig. \ref{fig:Sw_a}(a)]. The inset further shows that all the data with varying coupling strengths collapse on to one curve, implying emergence of a universal scaling whose analytical form will be given below. In the strong coupling regime, $S(\omega)$ simply follows a white noise spectrum over the entire frequency range [Fig. \ref{fig:Sw_a}(b)].  
%We admit that it is beyond our ability to analytically provide a comprehensive picture in the whole coupling regime. 
In order to understand such distinct behaviors, we note that the NE-PTRE reduces to the conventional quantum Redfield master equation (RME) and nonequilibrium non-interacting blip approximation (NE-NIBA) in the weak and strong coupling regime, respectively \cite{Wang.15.SR,Xu.16.NJP,Wang.17.PRA}. Thus we focus on these two limits and present analytical analyses in the following.

\subsubsection{Weak coupling regime}
\begin{figure}[tbh!]
  \centering
  \includegraphics[width=0.9\columnwidth]{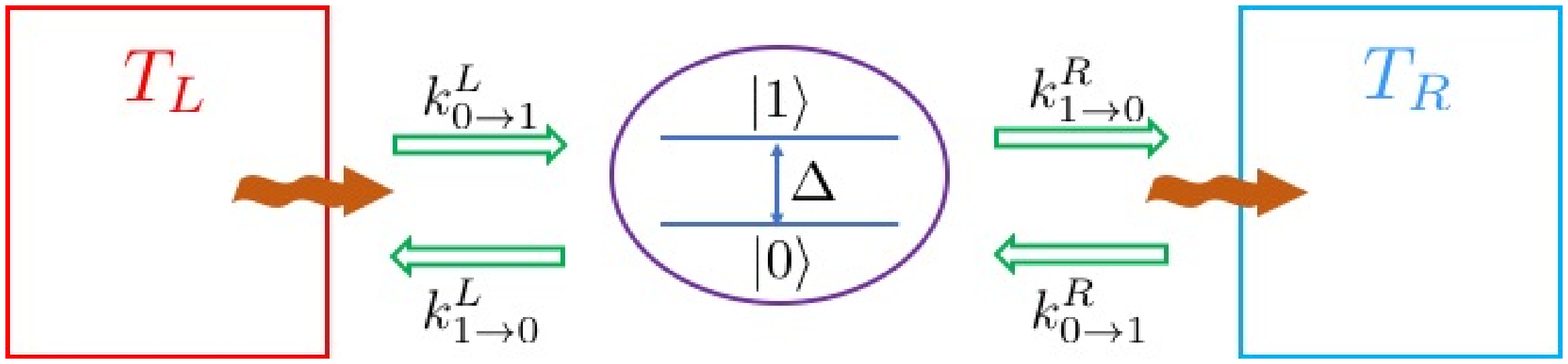}%0.9
\caption{(Color online) Schematic picture for the spin-boson model in the Redfield master equation framework.}
\label{fig:rate}
\end{figure}

We firstly concentrate on the weak coupling case and consider RME for the reduced dynamics \cite{Segal.06.PRB,Ren.10.PRL} (see the schematic picture in Fig. \ref{fig:rate}). 
We denote the relaxation and activation rates due to the $v$-th reservoir as
\begin{equation}\label{eq:kv}
k_{1\to 0}^v=\gamma_v(\Delta)[1+n_v(\Delta)],~~k_{0\to 1}^v~\equiv~k_{1\to0}^ve^{-\beta_v\Delta},
\end{equation}
respectively, where $\gamma_v$ and $n_v$ are the spectral density and Bose-Einstein distribution of $v$-th bath, respectively. Introducing $p_n(n=0,1)$ as the probability of the spin system to cccupy the state $|n\rangle$, satisfying $p_0(t)+p_1(t)=1$, then we have
\begin{equation}\label{eq:rho_weak}
|\dot{\rho}_s(t)\rangle\rangle~=~-
\left(
\begin{array}{cc}
k_d & -k_u\\
-k_d & k_u
\end{array}
\right)|\rho_s(t)\rangle\rangle
=-\mathbb{L}^R|\rho_s(t)\rangle\rangle,
\end{equation}
where $|\rho_s(t)\rangle\rangle=(p_1,p_0)^{T}$ and the total activation and relaxation rates read
\begin{equation}
k_u~=~\sum_vk_{0\to 1}^v,~~~k_d~=~\sum_vk_{1\to 0}^v,
\end{equation}
respectively. From the above rate equation, the stationary state solution corresponds to $|\rho_{s}^{stat}\rangle\rangle=\frac{1}{k_d+k_u}(k_u, k_d)^{T}$ which is just the right zero-eigenvector of $\mathbb{L}^R$, and the corresponding left zero-eigenvector reads $\langle\langle \tilde{0}|=(1,1)$ such that $\langle\langle \tilde{0}|\rho_{s}^{stat}\rangle\rangle=1$.

To study the statistics of heat, we split $p(Q,t)$ [Eq. (\ref{eq:pqt})] into two part, namely, $p(Q,t)=p_0(Q,t)+p_1(Q,t)$, where $p_0(Q,t)$ ($p_1(Q,t)$) denotes the probability that having $Q$ net heat transferred from the left reservoir into the right reservoir, within time interval $[0,t]$, while the spin is dwelling on the $|0\rangle$ ($|1\rangle$) energy level at time $t$ (the counting begins at $t=0$). By applying the transformation $p_n(\chi,t)=\int dQp_n(Q,t)e^{i\chi Q}$, we find
\begin{equation}\label{eq:px_weak}
|\dot\rho_s(\chi,t)\rangle\rangle~=~-
\left(
\begin{array}{cc}
k_d & -\tilde{k}_u\\
-\tilde{k}_d & ku
\end{array}
\right)|\rho_s(\chi,t)\rangle\rangle
=-\mathbb{L}_{\chi}^R|\rho_s(\chi,t)\rangle\rangle
\end{equation}
with $|\rho_s(\chi,t)\rangle\rangle=(p_1(\chi,t),p_0(\chi,t))^T$, $\tilde{k}_d=k_{1\to 0}^Le^{i\chi\Delta}+k_{1\to 0}^R$ and $\tilde{k}_u=k_{0\to 1}^Le^{-i\chi\Delta}+k_{0\to 1}^R$. A cumbersome evaluation within the Redfield picture yields the following, nontrivial explicit expression for $S(\omega)$ valid in the weak coupling regime
\begin{equation}\label{eq:sw_weak}
\frac{S(\omega)}{\Delta^2}~=~R-\frac{2k_L^2}{[D(\omega^2+D^2)]}[D^2e^{-\beta_L\Delta}+(k_u-k_de^{-\beta_L\Delta})^2],
\end{equation}
where $D=k_d+k_u$ is the sum of total relaxation and activation rates, $R=(k_uk_L+k_dk_Le^{-\beta_L\Delta})/D=p_1^{ss}k_{1\to 0}^L+p_0^{ss}k_{0\to 1}^L$ is the dynamical activity \cite{Maes.06.PRL,Lecomte.07.JSP}, which is the average number of transitions per time induced by the left reservoir. From the above equation, we see that $S(\omega)$ increases from $S(0)$ as $\omega$ increases and finally saturates at the value determined by the dynamical activity $R$, in accordance with numerical results shown in Fig. \ref{fig:Sw_a} (a). If we define a scaled \FD{}
\begin{equation}\label{eq:ssw}
\tilde{S}(\omega)~\equiv~R\Delta^2-S(\omega),
\end{equation}
a direct consequence of Eq. (\ref{eq:sw_weak}) is that $\tilde{S}(\omega)$ has a universal scaling expression
\begin{equation}\label{eq:uni_weak}
\frac{\tilde{S}(\omega)}{\tilde{S}(0)}~=~\mathcal{P}(\omega/D)
\end{equation}
with the scaling function $\mathcal{P}$ endows a Lorentzian shape and approaches $1$ as $\omega\to 0$. This universal behavior is manifested in our numerical results as can be seen from the inset in Fig. \ref{fig:Sw_a}(a). 

It would be interesting to see whether a similar scaling form in the weak coupling regime holds beyond the NESB model. For multi-level systems, the rates are still proportional to the coupling strength $\alpha$ in the weak coupling regime, we then expect a scaling function $\mathcal{F}(\omega/\alpha)$ still exists, however, the existence of multiple time scales will result in a complicated functional form of $\mathcal{F}$. Only for systems with a single time scale as Eq. (\ref{eq:sw_weak}) shows, the function $\mathcal{F}$ endows a Lorentzian shape. In future works, we also desire to look at universal behaviors of time dependent current noise as it is proportional to the variance of phonon numbers involved in the heat transfer in the weak coupling regime, the latter can be studied by a time dependent Poisson indicator \cite{Cao.08.JPCB}.

\subsubsection{Strong coupling regime}
We now turn to the white noise spectrum in the strong coupling regime, where the NE-PTRE is consistent with the NE-NIBA framework \cite{Wang.15.SR,Wang.17.PRA}. Using the NE-NIBA, the population dynamics with zero bias satisfies \cite{Nicolin.11.JCP,Nicolin.11.PRB,Chen.13.PRB}
\begin{equation}\label{eq:eom0}
|\dot{\rho}_s(t)\rangle\rangle~=~-\left(
\begin{array}{cc}
K & -K\\
-K & K
\end{array}
\right)|\rho_s(t)\rangle\rangle=-\mathbb{L}^N|\rho_s(t)\rangle\rangle,
\end{equation}
where the transition rate $K$ is given by \cite{Wang.15.SR,Wang.17.PRA}
\begin{equation}
K~=~(\eta\Delta/2)^2\int_{-\infty}^{\infty}\,dt e^{Q_L(t)+Q_R(t)}.
\end{equation}
From the equation of motion, the stationary state can be obtained as
\begin{equation}
\langle\langle\tilde{0}|~=~\frac{1}{2}(1,1),~~~|\rho_{s}^{stat}\rangle\rangle~=~(1,1)^T.
\end{equation}
In contrast to Eq. (\ref{eq:rho_weak}) of RME, now the diagonal and off-diagonal elements of $\mathbb{L}^N$ equal separately. As can be seen in the following, this distinctive spin dynamics with a single transition rate leads to the white noise we observed in Fig. \ref{fig:Sw_a}(b).

By incorporating the counting field, the equation of motion Eq. (\ref{eq:eom0}) becomes
\begin{eqnarray}\label{eq:eom_chi}
|\dot{\rho}_s(\chi,t)\rangle\rangle &=& -\left(
\begin{array}{cc}
K & -K_{\chi}\\
-K_{\chi} & K
\end{array}
\right)|\rho_s(\chi,t)\rangle\rangle\nonumber\\
&=&-\mathbb{L}_{\chi}^N|\rho_s(\chi,t)\rangle\rangle
\end{eqnarray}
with the $\chi$-dependent transition rate $K_{\chi}~=~(\eta\Delta/2)^2\int_{-\infty}^{\infty}\,dt e^{Q_L(t-\chi)+Q_R(t)}$  \cite{Wang.15.SR,Wang.17.PRA}. According to Eq. (\ref{eq:sw_SB}), after some algebras, we find
\begin{equation}
S(\omega) ~=~ \left.\frac{\partial^2}{\partial(i\chi)^2}K_{\chi}\right|_{\chi=0},
\end{equation}
which is just $S(0)$ by definition, thus we demonstrate that $S(\omega)$ is indeed a white noise spectrum and confirms our finding in the strong coupling regime as Fig. \ref{fig:Sw_a}(b) shows.

To gain more insights, we look at the explicit expression for the MGF. By diagonalizing the matrix $\mathbb{L}_{\chi}^N$, we find eigenvalues $\mu_0(\chi)=K-K_{\chi}$ and $\mu_1(\chi)=K+K_{\chi}$, the corresponding eigenvectors read $\langle\langle g_n(\chi)|=\frac{(K_{\chi},K-\mu_n(\chi))}{K_{\chi}^2+(K-\mu_n(\chi))^2}$ and $|f_n(\chi)\rangle\rangle= (K_{\chi},K-\mu_n(\chi))^T$ with $n=0,1$. It is evident that $\langle\langle g_1(\chi)|\rho_{s}^{stat}\rangle\rangle=\langle\langle\tilde{0}|f_1(\chi)\rangle\rangle=0$, then we find from Eq. (\ref{eq:z_e}) that
\begin{equation}
Z(\chi,t)~=~e^{-\mu_0(\chi)t}
\end{equation}
which is exactly the MGF obtained in the infinite time limit \cite{Nicolin.11.JCP,Chen.13.PRB}, thus we should have $S(\omega)=S(0)$ in this parameter regime.

We remark that a single transition rate in the population dynamics means the activation and relaxation rates equal, which is only possible in the high temperature regime as those bath-specific rates satisfy the detailed balance relation \cite{Nicolin.11.JCP}. In this regime, the memory of the system is totally destroyed by environments. Therefore, we find white noise spectrum for the \FD{} in the NESB model. It is desirable to investigate the \FD{} in systems consisting of multi-states, for such setups, interference effects plays an important role in transition rates at strong system-bath couplings \cite{Jang.01.JCP} which may change the behaviors of the \FD{}.

\subsection{Effect of bias}
\begin{figure}[tbh!]
  \centering
  \includegraphics[width=1\columnwidth]{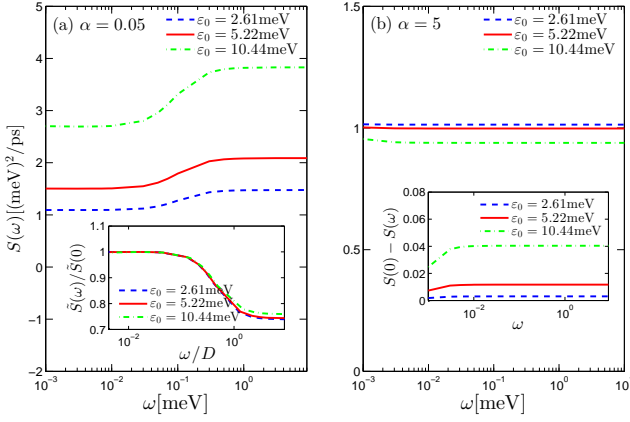}
\caption{(Color online) Behaviors of noise spectrum $S(\omega)$ as a function of $\omega$ with varying bias $\varepsilon_0$ for (a) $\alpha=0.05$ and (b) $\alpha=5$. The inset in (a) shows universal behaviors of a scaled noise power $\tilde{S}(\omega)$ defined in Eq. (\ref{eq:ssw}) with scale parameters given by Eq. (\ref{eq:sp_bias}) in the weak coupling regime, inset in (b) presents the deviation of $S(\omega)$ from $S(0)$ as a function of $\omega$. Other parameters are $\Delta=5.22$meV, $\omega_c=26.1$meV,$T_L=180$K, $T_R=90$K.}
\label{fig:Sw_bias}
\end{figure}

In the presence of bias, we still focus on the two coupling strength limits. The numerical results based on the $\chi$-dependent NE-PTRE [Eq. (\ref{eq:ptre_chi})] are shown in Fig. \ref{fig:Sw_bias}. 
In the weak coupling regime with $\alpha=0.05$ [Fig. \ref{fig:Sw_bias} (a)], behaviors of $S(\omega)$ as a function of $\omega$ with nonzero bias are similar to those with zero bias in Fig. \ref{fig:Sw_a} (a) and the \FD{} increases as the bias increases, implying that fluctuations are more prominent with larger bias in this regime. 

For weak couplings, we can use the energy basis of the two level system. Nonzero bias will change the energy gap from $\Delta$ to $\omega_0$ with the system Hamiltonian reads $H_s=\omega_0\sigma_z/2$. The original interaction term becomes
\begin{equation}
H_I~=~\sum_v(\sigma_z\cos\theta-\sigma_x\sin\theta)\otimes B_v
\end{equation}
with $B_v$ given by Eq. (\ref{eq:bv}) and $\theta=\tan^{-1}(\Delta/\omega_0)$. It is evident that only the $\sigma_x$ component in the interaction term contributes to spin-flip processes and thus to heat transfer in the Redfield picture. This implies that the transition rate $k_v$ defined in Eq. (\ref{eq:kv}) should be replaced by $\sin^2\theta k_v$ in the presence of nonzero bias \cite{Boudjada.14.JPCA}. Therefore, if we make the following replacements in Eqs. (\ref{eq:sw_weak}) and (\ref{eq:ssw})
\begin{equation}\label{eq:sp_bias}
\Delta\to\omega_0,~~~R\to \sin^2\theta R,~~~D\to\sin^2\theta D,
\end{equation}
then the universal relation Eq. (\ref{eq:uni_weak}) can still be applied to nonzero bias situations, as confirmed by our numerical results presented in the inset of Fig. \ref{fig:Sw_bias} (a). 

However, for strong couplings, nonzero bias leads to totally distinct behaviors compared with the zero bias case. As can be seen from the inset of Fig. \ref{fig:Sw_bias} (b), now $S(\omega)$ is no longer a white noise spectrum and suppressed by the bias, in direct contrast to its zero frequency counterpart which is insensitive to the bias change \cite{Wang.17.PRA}. To understand the role of finite bias in the strong coupling limit, we note the $\chi$-dependent Liouvillian operator in the NE-NIBA framework now becomes \cite{Nicolin.11.JCP,Nicolin.11.PRB,Chen.13.PRB}
\begin{equation}
\mathbb{L}_{\chi}^N~=~\left(\begin{array}{cc}
K(\varepsilon_0) & -K_{\chi}(-\varepsilon_0)\\
-K_{\chi}(\varepsilon_0) & K(-\varepsilon_0)
\end{array}
\right),
\end{equation}
where $K_{\chi}(\pm \varepsilon_0)=(\eta\Delta/2)^2\int_{-\infty}^{\infty}\,dt e^{\pm i\varepsilon_0 t+Q_L(t-\chi)+Q_R(t)}$ and $K(\pm \varepsilon_0)=\left.K_{\chi}(\pm \varepsilon_0)\right|_{\chi=0}$ are transfer rates.

By diagonalizing $\mathbb{L}_{\chi}^N$, we find $\mu_0(\chi) =\frac{1}{2}\left[\Xi(\varepsilon_0)-\Xi_{\chi}(\varepsilon_0)\right]$ and $\mu_1(\chi)=\frac{1}{2}\left[\Xi(\varepsilon_0)+\Xi_{\chi}(\varepsilon_0)\right]$, where we denote $\Xi(\varepsilon_0)\equiv K(\varepsilon_0)+K(-\varepsilon_0)$ and $\Xi_{\chi}(\varepsilon_0)\equiv\sqrt{(K(\varepsilon_0)-K(-\varepsilon_0))^2+4K_{\chi}(-\varepsilon_0)K_{\chi}(\varepsilon_0)}$, the corresponding eigenvectors read
\begin{eqnarray}
\langle\langle g_n(\chi)| &=& \frac{(K_{\chi}(\varepsilon_0),K(\varepsilon_0)-\mu_n(\chi))}{K_{\chi}(-\varepsilon_0)K_{\chi}(\varepsilon_0)+(K(\varepsilon_0)-\mu_n(\chi))^2},\nonumber\\
|f_n(\chi)\rangle\rangle &= & (K_{\chi}(-\varepsilon_0),K(\varepsilon_0)-\mu_n(\chi))^T.
\end{eqnarray}
Since $\langle\langle g_1(\chi)|\rho_{s}^{stat}\rangle\rangle\neq0$ and $\langle\langle\tilde{0}|f_1(\chi)\rangle\rangle\neq0$, the resulting MGF obviously no longer equals $e^{-\mu_0(\chi)t}$ as it contains a contribution from the eigenvalue $\mu_1(\chi)$ according to Eq. (\ref{eq:z_e}), thus we expect frequency dependence of $S(\omega)$ in the presence of bias as Fig. \ref{fig:Sw_bias} (b) shows. 

\subsection{Effect of temperature difference}
Now we extend our analysis of \FD{} to the impact of temperature difference ranging from the linear response regime to nonlinear situation. 

\subsubsection{$\omega=0$: Thermodynamic consistency}
In the zero frequency limit, the NESB model 
we consider satisfies the Gallavotti-Cohen (GC) symmetry \cite{Gallavotti.95.PRL} as shown in previous studies \cite{Ren.10.PRL,Nicolin.11.PRB}, thus the Saito-Utsumi (SU) relations can be applied to $J_m^n=\lim_{t\to\infty}\frac{\partial}{\partial t}C_m^n(t)$ [see Eq. (\ref{eq:gsu})], yielding \cite{Saito.08.PRB} 
\begin{equation}\label{eq:su}
J_m^n~=~\sum_{l=0}^m\left(
\begin{array}{c}
m\\
l
\end{array}
\right)(-1)^{n+l}J_{m-l}^{n+l},
\end{equation}
from which we find $2J_1^1=J_0^2$ and thus $S(0)=2\frac{\partial I_L}{\partial A}$ by noting $S(0)=J_0^2$.
Associated with the coefficient $J_1^1$, we can introduce a first-order energetic transport coefficient as \cite{Cerrillo.16.PRB}: $\kappa_F\equiv\beta_L\beta_RJ_1^1$, therefore \begin{equation}\label{eq:s01}
S(0)~=~2T_LT_R\kappa_F.
\end{equation}
For small temperature differences, Eq. (\ref{eq:s01}) reduces to a linear response relation \cite{Averin.10.PRL}
\begin{equation}\label{eq:lr}
S(0)~=~2T^2\kappa
\end{equation}
with $\kappa$ the heat conductance. For later purpose, first we check that our theory indeed satisfies Eq. (\ref{eq:s01}) and thus preserves the GC symmetry in the zero frequency limit. 
\begin{figure}[tbh!]
  \centering
  \includegraphics[width=1\columnwidth]{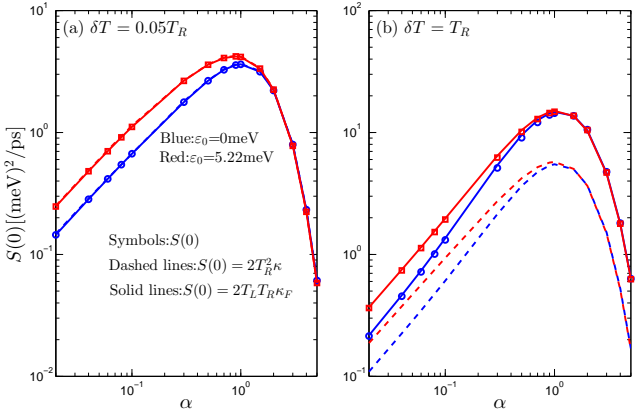}
\caption{(Color online) Behaviors of noise spectrum $S(0)$ as a function of $\alpha$ with varying bias for (a) temperature difference $\delta T=0.05T_R$ and (b) $\delta T=T_R$. Symbols are direct results of $S(0)$ using our theory, dashed lines are predictions of Eq. (\ref{eq:lr}), solid lines are predictions of Eq. (\ref{eq:s01}). Other parameters are $\Delta=5.22$meV, $\omega_c=26.1$meV,$T_L=T_R+\delta T$, $T_R=90$K.}
\label{fig:FDR}
\end{figure}
As shown in Fig. \ref{fig:FDR}, we clearly see good agreements between numerical results and theoretical relations. Eq. (\ref{eq:lr}) captures the behaviors of $S(0)$ with small temperature differences in the entire coupling strength range regardless of values of bias, while Eq. (\ref{eq:s01}) holds generally in our theory regardless of the magnitude of temperature difference. 

\subsubsection{Frequency dependence}
\begin{figure}[tbh!]
  \centering
  \includegraphics[width=1\columnwidth]{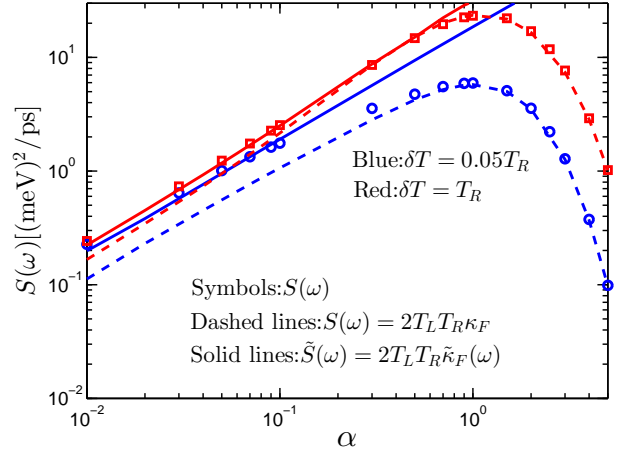}
\caption{(Color online) Behaviors of noise spectrum $S(\omega)$ as a function of $\alpha$ with varying temperature differences and fixed $\omega=0.5$. Blue color denotes $\delta T=0.05T_R$,  Red color denotes $\delta T=T_R$. Symbols are direct numerical results of $S(\omega)$ based on the $\chi$-dependent NE-PTRE Eq. (\ref{eq:ptre_chi}), dashed lines are  prediction of strong coupling relation Eq. (\ref{eq:s_strong}), solid lines are predictions of weak coupling expression Eq. (\ref{eq:s_weak}). Other parameters are $\Delta=5.22$meV, $\omega_c=26.1$meV, $\varepsilon_0=0$meV. $T_L=T_R+\delta T$, $T_R=90$K.}
\label{fig:Sw_T}
\end{figure}

Now we investigate $S(\omega)$. So far, there are no general relations between $S(\omega)$ and first order quantities characterizing the response to arbitrary temperature difference for finite frequency cases \cite{Averin.10.PRL}. However, according to above universal behaviors in the two coupling strength limits, we can formulate general relations valid in the corresponding coupling strength regimes. The white noise behavior in the strong coupling regime for unbiased systems implies that the SU relation Eq. (\ref{eq:s01}) can be directly applied to the \FD{}, namely,
\begin{equation}\label{eq:s_strong}
S(\omega)~=~2T_LT_R\kappa_F.
\end{equation}
While in the weak coupling regime, the universal scaling form Eq. (\ref{eq:uni_weak}) guarantees the following relation
\begin{equation}\label{eq:s_weak}
\tilde{S}(\omega)~=~2T_LT_R\tilde{\kappa}_F(\omega)
\end{equation}
for the scaled \FD{} $\tilde{S}(\omega)$ and a frequency-dependent first order coefficient $\tilde{\kappa}_F(\omega)=\mathcal{P}(\omega/D)\left[\kappa_F-\frac{\Delta^2R}{2T_LT_R}\right]$.  Their validity can be seen from comparisons in Fig. \ref{fig:Sw_T}. 
We find that the weak coupling expression Eq. (\ref{eq:s_weak}) predicts a monotonic increasing behavior of $S(\omega)$ as a function of $\alpha$, thus becomes invalid in the intermediate as well as strong coupling regimes. The strong coupling expression Eq. (\ref{eq:s_strong}) underestimates the current fluctuations in the weak coupling regime.  Although our theory can provide a detailed description for the \FD{} with arbitrary temperature differences in the two limits of coupling strength,  a general yet simple relation for $S(\omega)$ and temperature differences beyond these two coupling limits is desirable and will be addressed in the future works.

\section{Summary}\label{sec:5}
We formulate a general theory to study frequency-dependent current noise (\FD) in open quantum systems at steady states. To go beyond previous results, we extend MacDonald's formula from electron transport to heat current, and obtain a formally exact relation which relates the \FD{} to the time-dependent second order cumulant of heat evaluated at steady states. In order to calculate the time-dependent cumulant of heat involved in the \FD{}, we follow the scheme of a finite time full counting statistics (FCS) developed for electron transport and propose an analogous framework, which can be applied to open quantum systems described by Markovian quantum master equations.

To demonstrate the utility of the approach, we consider the nonequilibrium spin-boson model which is a paradigmatic example of quantum heat transfer.  A recently developed polaron-transformed Redfield equation for the reduced spin dynamics enables us to study the \FD{} from weak to strong system-reservoir coupling regimes and consider arbitrary values of bias and temperature differences. Key findings are: 

(1) By varying coupling strengths, we observe a turn-over behavior for the \FD{} in moderate coupling regimes, similar to the zero frequency counterpart. Interestingly, the \FD{} with varying coupling strength or bias exhibits a universal Lorentzian-shape scaling form in the weak coupling regime as confirmed by numerical results as well as analytical analysis, while it becomes a white noise spectrum under the condition of strong coupling strengths and zero bias. The white noise spectrum is distorted in the presence of a finite bias. 

(2) We also find the bias can suppress frequency-dependent current fluctuations in the strong coupling regime, in direct contrast to the zero frequency counterpart which is insensitive to the bias changes. 

(3) We further utilize the Saito-Utsumi (SU) relation as a benchmark to evaluate the theory at zero frequency limit in the entire coupling range. Agreements between SU relation and our zero frequency results shows that our theory preserves the Gallavotti-Cohen symmetry. Noting the universal behaviors of the \FD{} in the weak as well as strong coupling regime, we then study the impact of temperature differences at finite frequencies by carefully generalizing the SU relations. Our results thus provide detailed dissections and a unified framework for studying the finite time fluctuation of heat in  open quantum systems.

\begin{acknowledgments}
J. Liu thanks Chen Wang for correspondences on related topics and Michael Moskalets for interesting comments. The authors acknowledge the support from the Singapore-MIT Alliance for Research and Technology (SMART).
\end{acknowledgments}

%\bibliography{a}

\begin{thebibliography}{60}
\bibitem{Lee.13.N}W. Lee, K. Kim, W. Jeong, L. A. Zotti, F. Pauly, J. C. Cuevas, and P. Reddy, Nature {\bf498}, 209 (2013).
\bibitem{Jezouin.13.S}S. Jezouin, F. D. Parmentier, A. Anthore, U. Gennser, A. Ca- vanna, Y. Jin, and F. Pierre, Science {\bf342}, 601 (2013).
\bibitem{Battista.13.PRL}F. Battista, M. Moskalets, M. Albert, and P. Samuelsson, Phys. Rev. Lett. {\bf110}, 126602 (2013).
\bibitem{Touchette.09.PR}H. Touchette, Phys. Rep. {\bf478}, 1 (2009).
\bibitem{Jarzynski.04.PRL}C. Jarzynski and D. K. W\'ojcik, Phys. Rev. Lett. {\bf92}, 230602 (2004).
\bibitem{Esposito.09.RMP}M. Esposito, U. Harbola, and S. Mukamel, Rev. Mod. Phys. {\bf81}, 1665 (2009).
\bibitem{Seifert.12.RPP}U. Seifert, Rep. Prog. Phys. {\bf75}, 126001 (2012).
\bibitem{Levitov.93.PZ}L. S. Levitov and G. B. Lesovik, Pis?ma Zh. Eksp. Teor. Fiz {\bf58}, 225 (1993).
\bibitem{Bagrets.03.PRB}D. A. Bagrets and Y. V. Nazarov, Phys. Rev. B {\bf67}, 085316 (2003).
\bibitem{Agarwalla.12.PRE}B. K. Agarwalla, B. Li, and J.-S. Wang, Phys. Rev. E {\bf85}, 051142 (2012).
\bibitem{Boudjada.14.JPCA}N. Boudjada and D. Segal, J. Phys. Chem. A {\bf118}, 11323 (2014).
\bibitem{Wang.15.SR}C. Wang, J. Ren, and J. Cao, Sci. Rep. {\bf5}, 11787 (2015).
\bibitem{Liu.17.PRE}J. Liu, H. Xu, B. Li, and C. Wu, Phys. Rev. E {\bf96}, 012135 (2017).
\bibitem{Saito.07.PRL}K. Saito and A. Dhar, Phys. Rev. Lett. {\bf99}, 180601 (2007).
\bibitem{Nicolin.11.JCP}L. Nicolin and D. Segal, J. Chem. Phys. {\bf135}, 164106 (2011). 
\bibitem{Nicolin.11.PRB}L. Nicolin and D. Segal, Phys. Rev. B {\bf84}, 161414 (2011).
\bibitem{Wang.17.PRA}C. Wang, J. Ren, and J. Cao, Phys. Rev. A {\bf95}, 023610 (2017). 
\bibitem{Bijay.17.NJP}B. K. Agarwalla and D. Segal, New J. Phys. {\bf19}, 043030 (2017). 
\bibitem{Zhan.11.PRB}F. Zhan, S. Denisov, and P. H\"anggi, Phys. Rev. B {\bf84}, 195117 (2011).
\bibitem{Krive.01.PRB}I. V. Krive, E. N. Bogachek, A. G. Scherbakov, and U. Landman, Phys. Rev. B {\bf64}, 233304 (2001).
\bibitem{Averin.10.PRL}D. V. Averin and J. P. Pekola, Phys. Rev. Lett. {\bf104}, 220601 (2010).
\bibitem{Breuer.07.NULL}H.-P. Breuer and F. Petruccione, {\it The Theory of Open Quantum Systems} (Oxford University Press, Oxford, 2007).
\bibitem{Weiss.12.NULL}U. Weiss, {\it Quantum Dissipative Systems} (World Scientific, Singapore, 2012).
\bibitem{Schaller.14.NULL}G. Schaller, {\it Open Quantum Systems Far From Equilibrium} (Springer, Heidelberg, 2014).
\bibitem{MacDonald.49.RPP}D. K. C. MacDonald, Rep. Prog. Phys {\bf12}, 56 (1949).
\bibitem{Lambert.07.PRB}N. Lambert, R. Aguado, and T. Brandes, Phys. Rev. B {\bf75}, 045340 (2007).
\bibitem{Marcos.10.NJP}D. Marcos, C. Emary, T. Brandes, and R. Aguado, New J. Phys. {\bf12}, 123009 (2010).
\bibitem{Cerrillo.16.PRB}J. Cerrillo, M. Buser, and T. Brandes, Phys. Rev. B {\bf94}, 214308 (2016).
\bibitem{Leggett.87.RMP}A. J. Leggett, S. Chakravarty, A. T. Dorsey, M. P. A. Fisher, A. Garg, and W. Zwerger, Rev. Mod. Phys. {\bf59}, 1 (1987).
\bibitem{Wang.14.NJP}C. Wang, J. Ren, and J. Cao, New J. Phys. {\bf16}, 045019 (2014).
\bibitem{Xu.16.NJP}D. Xu, C. Wang, Y. Zhao, and J. Cao, New J. Phys. {\bf18}, 023003 (2016).
\bibitem{Thingna.16.SR}J. Thingna, D. Manzano, and J. Cao, Sci. Rep. {\bf6}, 28027 (2016).
\bibitem{Esposito.15.PRL}M. Esposito, M. A. Ochoa, and M. Galperin, Phys. Rev. Lett. {\bf114}, 080602 (2015).
\bibitem{Flindt.05.PE}C. Flindt, T. Novotn\'y, and A.-P. Jauho, Physica E (Amsterdam) {\bf29}, 411 (2005).
\bibitem{Campisi.11.RMP}M. Campisi, P. H\"anggi, and P. Talkner, Rev. Mod. Phys. {\bf83}, 771 (2011).
\bibitem{Braggio.06.PRL}A. Braggio, J. Koenig and R. Fazio, Phys. Rev. Lett. {\bf96}, 026805 (2006).
\bibitem{Flindt.08.PRL}C. Flindt, T. Novotn\'y, A. Braggio, M. Sassetti, and A.-P. Jauho, Phys. Rev. Lett. {\bf100}, 150601 (2008).
\bibitem{Cook.81.PRA}R. J. Cook, Phys. Rev. A {\bf23}, 1243 (1981).
\bibitem{Gurvitz.98.PRB}S. A. Gurvitz, Phys. Rev. B {\bf57}, 6602 (1998).
\bibitem{Goerlich.89.EL}R. G\"orlich, M. Sassetti, and U. Weiss, Europhys. Lett. {\bf10}, 507 (1989).
\bibitem{Lee.12.JCP}C. K. Lee, J. Moix, and J. Cao, J. Chem. Phys. {\bf136}, 204120 (2012).
\bibitem{Liu.16.CP}J. Liu, H. Xu, and C. Wu, Chem. Phys. {\bf481}, 42 (2016).
\bibitem{Segal.06.PRB}D. Segal, Phys. Rev. B {\bf73}, 205415 (2006).
\bibitem{Chen.13.PRB}T. Chen, X.-B. Wang, and J. Ren, Phys. Rev. B {\bf87}, 144303 (2013).
\bibitem{Ren.10.PRL}J. Ren, P. H\"anggi, and B. Li, Phys. Rev. Lett. {\bf104}, 170601 (2010).
\bibitem{Maes.06.PRL}C. Maes and M. H. van Wieren, Phys. Rev. Lett. {\bf96}, 240601 (2006).
\bibitem{Lecomte.07.JSP}V. Lecomte, C. Appert-Rolland, and F. van Willand, J. Stat. Phys. {\bf127}, 51 (2007).
\bibitem{Cao.08.JPCB}J. Cao and R. J. Silbey, J. Phys. Chem. B {\bf112}, 12867 (2008).
\bibitem{Jang.01.JCP}S. Jang and J. Cao, J. Chem. Phys. {\bf114}, 9959 (2001).
\bibitem{Gallavotti.95.PRL}G. Gallavotti and E. G. D. Cohen, Phys. Rev. Lett. {\bf74}, 2694 (1995).
\bibitem{Saito.08.PRB}K. Saito and Y. Utsumi, Phys. Rev. B {\bf78}, 115429 (2008).
\end{thebibliography}

\end{document}